# Resource Management of energy-aware Cognitive Radio Networks and cloud-based Infrastructures

Evangelos Katis
Department of Informatics Engineering
Technological Educational Institute of Crete
Heraklion, Greece
evskg@sch.gr

*Abstract* - **The field of wireless networks has been rapidly developed during the past decade due to the increasing popularity of the mobile devices. The great demand for mobility and connectivity makes wireless networking a field whose continuous technological development is very important as new challenges and issues are arising. Many scientists and researchers are currently engaged in developing new approaches and optimization methods in several topics of wireless networking. This survey paper study works from the following topics:** *Cognitive Radio Networks, Interactive Broadcasting, Energy Efficient Networks, Cloud Computing and Resource Management, Interactive Marketing and Optimization.*

I. INTRODUCTION

The advent of mobility has raised a great demand in wireless connectivity and bandwidth. A growing number of mobile users along with bandwidth-hungry applications introduce new needs and issues for wireless networking technologies and new challenges need to be addressed in order to provide wireless services at guaranteed QoS. Wireless networks have to provide high bandwidth and quality services to mobile users via heterogeneous wireless architectures.

However, due to the fluctuating nature of wireless networks, there are plenty of issues in several topics that need investigation and optimization. In this survey paper, a lot of related works are studied and presented, following a categorized structure based on their topic:

- *Section II: Cognitive Radio Networks*˙ provide the necessary spectrum availability for high bandwidth wireless connectivity.
- *Section III: Interactive Broadcasting*˙ confronts the challenge of broadcasting digital TV programmes and IP-based triple-play services via a DVB-T channel.
- *Section IV: Energy-Efficient Networks*˙ offer the communication nodes Energy Conservation via Energy Consumption.
- *Section V: Cloud Computing and Resource Management*˙ introduces optimization methods for efficient resource sharing in a cloud computing environment.
- *Section VI: Interactive Marketing*˙ enables optimum marketing and advertising techniques by collecting and analyzing feedback data from digital tele-viewers.
- *Section VII: Optimization*˙ several optimization methods, algorithms and mechanisms are presented as important keys in wireless networking.

## II. COGNITIVE RADIO NETWORKS

Cognitive Radio Technology provides an emerging and promising communication model that exploits efficiently the resource management of radio spectrum, enabling the deployment of future sophisticated wireless networks.

Formally, a Cognitive Radio is defined as *a radio that can change its transmitter parameters based on interaction with its environment*.

The powerful benefit of Cognitive Radio (CR) Networks are the innovative cognitive radio techniques that provide the capability to share the spectrum in an opportunistic manner. Two main characteristics can be defined:

- *Cognitive capability:* Based on communication nodes capable of dynamically identify unused/unexploited frequencies through real-time interaction with the radio environment.
- *Reconfigurability*: Through this capability, they can be programmed to adjust their operation in order to select the best spectrum band with the most appropriate operating parameters [1].

Since CR Networks is maybe the most promising technology for future wireless networks, the available literature continues to be enriched. A variety of approaches in network implementations is available with several proposals on exploiting efficiently the unused radio spectrum, utilized by promising algorithms and protocols.

The superior transmission efficiency of Digital Video Broadcasting (DVB) results in a spectrum release that CR networks are employed to exploit. This spectrum release will be available when Digital Switchover is accomplished. *Digital Switchover (DSO) is the switchover move from analogue to digital television* [2].

TV White spaces (TVWS) comprise radio spectrum portions that are either liberalized as a result of DSO process or are completely unused due to Interleaved Spectrum.

*Interleaved Spectrum represents the amount of frequencies which remain unused between adjacent broadcast coverage areas* because, the channels used in one region cannot be used in adjacent regions in order to prevent interference [3]. Especially in rural areas, where some national TV channels are absent, the TVWS presented to be a perfectly suitable and economic solution for broadband access, large scale wi-fi , gaming, monitoring applications, etc [3], [4], [5].

Furthermore, by the appropriate adoption of DSO, new radio spectrum opportunities are coming up due to dynamic radio spectrum management capabilities and policies [2]. This process liberalizes valuable TV radio frequencies, either in the form of cleared spectrum of cleared analogue channels as well as in the form of TV White Spaces [6], introducing the digital dividend as a unique opportunity to gain economic and social benefits, leveraging entrepreneurship and boosting competitiveness through new business strategies especially of small business players in ICT sector [7]. The exploitation of Interleaved Spectrum arises the opportunity of new wireless network technologies and services such as local digital terrestrial television networks, mobile TV systems, wireless broadband network, public safety and cognitive radio systems, new generation mobile networks (LTE) [8]. Medium or small business players could provide such services, like always-on connectivity and triple-play services access. Especially in rural and less developed regions, with the population being normally un-served with broadband access, the delivery of digital TV bouquets and network services offers an enormous improvement of the economic and social life [9]. With regard to the benefits in economy, it is reported that the DVB services will generate over 700 bn € in net present value for the European economy [2].

Digital Dividend introduces new wireless network technologies and services. Dynamic Spectrum Access (DSA) is highly related to CR networks enabling exploitation of radio spectrum resources efficiently. Based on this, two main categories of network architectures can be defined [10]:

- *Centralized*: The decision of radio spectrum sharing and access is taken by a central controller.
- *Distributed*: Each individual node is capable of discovering unexploited frequencies and adjusting their operation parameters, without the presence of a central unit.

Depending on the network architecture, novel policies for spectrum administration and management are being introduced respectively with licensed and unlicensed policy.

Along with the centralized architecture is the "Real-time Secondary Spectrum Market (RTSSM)" where licensed policy enables primary systems to trade rights of networking resources with secondary systems. In this case, primary systems run network management algorithms, dealing with secondary systems for frequency access. Thus, a secondary market of spectrum is established for radio spectrum leasing. A competitive secondary market enables associates to deal the price of spectrum at the required quality of service [2], [5], [11]. An approach where the exploitation of the available TVWS, as well as the administration of the economic transactions is orchestrated via a Spectrum Broker as it is elaborated by A.Bourdena *et al.* [12], [13], [14], [15]. Furthermore, a broker-based architecture with QoS provisioning and policy management, is presented in Ref. [16], [17].

Turning now to the other approach, the "Spectrum of Commons" a regime is adopted in the distributed architecture with no spectrum manager and no fixed spectrum allocation policy but the decision of resource allocation is taked locally. In this implementation, the communication nodes are capable of cognitive interaction with other systems and dynamically reconfigured in order to avoid possible interferences by shifting to another available bandwidth area. This process guarantees a conflict free coexistence with other networks and elevates network efficiency through radio spectrum sharing. Sensing techniques for reliable detection of free frequencies and coexistence mechanisms for interference avoidance are the main technical challenge here.

A routing protocol for effective communication of secondary communication nodes with a signaling mechanism based on end-to-end bounded delay of the transmission is proposed in Ref. [1]. A resource intensive traffic-aware scheme, incorporated into an energy-efficient routing protocol that enables energy conservation and efficient data flow coordination, among secondary communicating nodes with heterogeneous spectrum availability is proposed in Ref. [18]. Moreover, a spectrum aware routing protocol that coordinates data flows and establishes optimum routing paths among secondary users with heterogeneous spectrum availability is discussed in Ref. [19]. This protocol is supposed to exploit the communication of the nodes both Ad-hoc and mesh network architectures. Ad-hoc network connections are ideal in case that the spectrum resources are pour, while the mesh network architecture addresses market's emerging requirements for building highly scalable and cost effective wireless networks [20]. E. Papadopoulos *et al.* [21] reports a performance analysis of reactive routing protocols in Mobile Ad hoc Networks, which are self-organized infrastructure-less networks where the nodes are able to move in any direction in the network showing unpredictable behavior.

It must also be noted that Quality of Service (QoS) is not possible to be guaranteed, which may become a critical issue, especially for QoS sensitive applications and applications [10], [17].

### III. INTERACTIVE BROADCASTING

Terrestrial digital video broadcasting standard (DVB-T) is commonly used for broadcasting digital TV programmes to a large number of viewers scattered over large geographical areas, urban and rural. The large coverage area, the high bit-rate capabilities and the intrinsic characteristic of DVB-T to combine MPEG-2 TV programmes and IP data traffic into a single transport stream, makes it an efficient and economic technology for usage as the "last mile" in networking infrastructures for the provision of digital TV programmes and IP-based services (i.e. Internet access, e-mail access, multimedia services on demand and/or in multicast form) [22].

Several solutions have been proposed for confronting this challenge, mainly based upon two approaches: the convergence telecommunication technologies and the Interactive Broadcasting. The cheap, easy and always-on access to the Information Society services (i.e. e-Business, e-Inclusion, e-Learning, e-Health, e-Government, etc.) is of major and strategic importance with evident results over the social, cultural and economic life of citizens, especially in less developed regions like rural and remote areas [23], [24].

ATHENA Concept is presented by E. Pallis *et all* [25], [26] as an approach for proper adoption of DSO that provides the basis for fast and successful establishment of broadband infrastructures. An environment that enables broadband access to triple play services via a DVB-T channel. Detailed figures also included to describe the overall architecture and the configuration of that fusion environment that comprises two core subsystems: a number of Cell Main Nodes (CMN) and a central broadcasting point (regenerative DVB-T). The users have access to the network services via broadband connections to the CMN which forward all users' IP traffic to the regenerative DVB-T. Furthermore, they demonstrate their proposal by building an infrastructure that multiplexes and transmits TV programs with IP-based services (i.e. Internet access, e-mail access, multimedia services on demand and/or in multicast form). The, so called, ATHENA infrastructure is capable to receive and distribute services and applications to the entire city of Heraklion via a DVB-T channel.

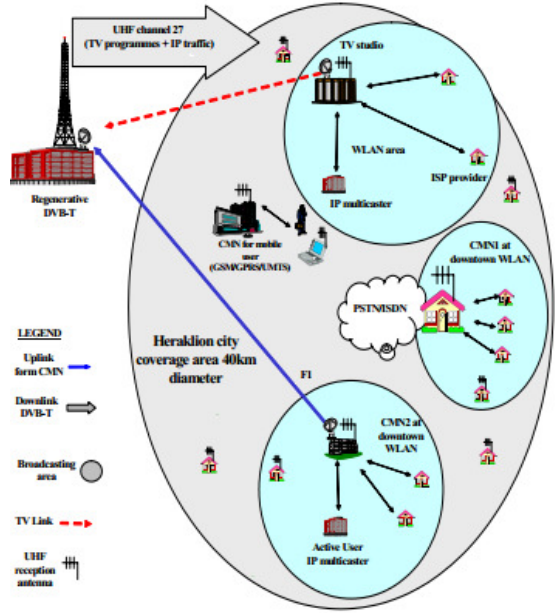

*Figure 1: Overall Architecture*

One of the most promising applications that could be supported are tele-learning applications such as the "virtual classroom" that enables lecturers, teachers and students to participate in the same classroom from their own residences [27].

In order to support such applications, QoS-based mechanisms are necessary. A DVB/IP backhaul environment that enables the broadcasting of triple-play services at guaranteed QoS requirements is proposed in Ref. [28], [29]. The designed system is validated under real transmission/reception conditions [9], [30].

Another ambitious application of DVB-T, is the Interactive Marketing. Since digital television systems can efficiently support collecting and analyzing feedback data from users/viewers, optimum marketing and advertising techniques may be used to focus on customer's preferences more successfully [31].

Security is always a significant issue in networks, so the importance to secure IP traffic conveyed over DVB-T channels must also be noted with a variety of techniques and protocols that can be used.

## IV. ENERGY-EFFICIENT NETWORKS

The design and operation scheme of networks and networks systems is a major issue because it is critical to support the expected traffic growth due to the increase of users and devices, the new services and new applications of future communication networks. Especially with the recent popularity of mobile devices, such as laptops, smartphones, tablets and wearable smartwatches, the need for reliable and high performance wireless network environments, is becoming even more crucial. As the awareness of the effects of unsustainable energy consumption increases, it is time to consider how networking can be made more efficient and green.

Several research efforts have introduced innovated approaches using different mechanisms to reduce energy consumption. They can be classified into two main categories [32]:

- *Active techniques* conserve energy by operating in more energy conscious modes, such as energy-aware routing, directional antennas usage and mobile cloud computing.
- *Passive techniques* conserve energy by setting the network interface to sleep mode when there is no communication activity.

A novel offloading-based energy conservation mechanism for wireless devices utilizes mobile cloud computing using a service that implements a cooperative partial process execution in an offloading scheme aiming at energy consumption due to lower processing requirements. It outsources, totally or partially, the required resources while running an application in order to minimize the CPU/GPU usage resulting in lowering the energy consumption. Such a mobile service has to be reliable, ensuring the effective execution of application under cooperative energy-efficient scheme. Therefore, a dynamic scheduling mechanism elaborated by a partial offloading algorithm offering an energy-efficient failure-aware allocation of the resources, to ensure that no discontinuous execution will happen on mobile devices.

An active technique uses a non-centric scheme with joint power and resource allocation methodology for energy-aware routing is suggested in Ref. [33]. Each node collaborates with other nodes, estimates the throughput and tunes the proposed path-aware energy conservation mechanism in order to provide the least power cost path. Furthermore, a routing protocol with the appropriate mechanism and a capacity-aware scheme is proposed by G.Mastorakis *et al.* [34], where networking nodes operate over TV White Spaces (TVWS). It aims to provide optimal, high throughput data transfer, by efficiently selecting the best routing paths [35]. Furthermore, in Ref. [36] two energy-aware algorithms are introduced, exploited for maximum energy conservation and reliability of secondary nodes during the resource sharing process in a centralized cognitive radio network. As for the administration of the resources trading process, there is a radio spectrum broker, based on real-time secondary spectrum policy.

A opportunistic networking paradigm related to passive techniques, recommended in Ref. [37], enables Energy Conservation (EC) by using Real-Time Backward Traffic Difference estimation *(Figure 3)*. The scheme schedules the next Sleep/Wake-time of the network interface in wireless devices considering the traffic's backward difference. A cached mechanism is monitoring the traffic that traverses the nodes and, by taking into consideration the repetition pattern of the traffic, it estimates the duration of the next sleep time of the node (one-level) and extracts the time duration for which the node is allowed to sleep (second-level) [32], [38].

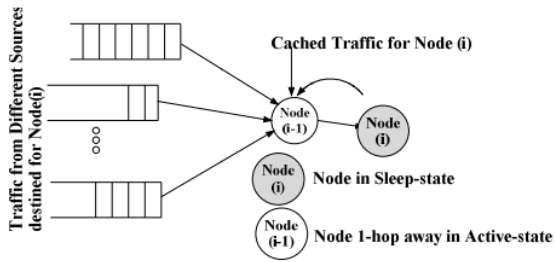

*Figure 2: A schematic diagram of the caching mechanism*

Thus, it reduces the Energy Consumption while at the same time it guarantees the end-to-end availability of requested resources. In addition, the Fibonacci-based Backward Traffic Difference (F-BTD) scheme [39], aims to prolong the lifetime of the nodes while optimizing the QoS and the overall throughput response of each node.

V. CLOUD COMPUTING AND RESOURCE MANAGEMENT

Cloud Computing is one of the emergence developments in the IT industry which significantly changed everyone's sense of infrastructure architectures, software delivery and development models. Compared to dedicated infrastructures, it provides a relatively low cost feasible solution with full scalability, reliability and high performance. However, despite the fact that cloud computing offers huge opportunities to the IT industry, the development of cloud computing technology is currently at its infancy, with many issues still to be addressed. Such issues as the relatively high operating cost for both public and private Clouds and the management of resources are becoming increasingly important under an ever-rising demand for more computational power and resources.

Software as a Service (SaaS) is a very popular paradigm in public clouds used by millions of users and applications. There is a great need to have the offered services provided to the end-users, with the minimum possible reduction in quality in terms of time and availability. Although bandwidth and access will continue to improve, the delay time is unlikely to get improved. Therefore, the main need in such a case is the deployment of a resource migration and offloading model. A cloud rack configuration is presented in Ref. [40] where nodes are allocated on cloud rack and resources are inter-exchanged among mobile and static users based on the best effort allocation.

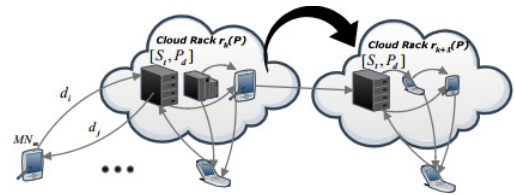

*Figure 3: Cloud Racks*

Nowadays, with the increasing popularity of mobile devices, such as laptops, smart phones, tablets and smart televisions, a huge number of applications is being developed to support them. Many of these applications require network connection and generate a lot of Internet traffic. A proper analytical description of the end-user traffic is described in Ref. [41] while measurements of modern wireless communication technologies and estimation of human exposure are presented. In case when no WiFi or Bluetooth connection is available, the only solution is the mobile broadband access which is not always free or unlimited. With modern applications demand for Internet access, subscribers have to be aware how they consume their broadband traffic.

OpenMobs is a proposed means for reducing the cost associated with mobile broadband Internet traffic by sharing under-utilized networking resources. An ad-hoc mesh network is formed by co-located users through free wireless connections for data transfer in the most economic and viable path. Furthermore, with social interaction in a mobile cloud, resources offloading schemes can be implemented in order to conserve energy without affecting the user's *Quality of Experience* (QoE) [42]. Energy consumption is important for wireless nodes since it affects the availability and reliability of the network. A cloud offload model using social centrality is analyzed. *Social Centrality* is the key feature in a cloud offload model where *the importance of a node based on its position, connectivity and interactivity patterns*. This model uses this social centrality aspect to decide which node is to be offloaded so as to minimize the total energy consumption providing a total higher availability for the most popular nodes [43]. Moreover, traffic aware mechanisms may be adopted in content-aware networks for efficient resource management [44], [45].

Another model for sharing information between mobile and vehicle devices is the Floating Content. The use of applications with location-aware mobile technology is the base for a decentralized architecture in order to achieve two of the main aspects of the smart cities: the *content sharing service* and the *vehicle-to-vehicle wireless communication.* Smart cities of tomorrow will use sensor networks to monitor and control applications but a critical issue of interoperability has to be addressed. For that issue, a solution is proposed in Ref. [46] that describes the provision of a unified service access architecture, which will support common interfaces for data communications.

## VI. INTERACTIVE MARKETING

*Interactive Marketing* is defined as *the trading situation where the buyer specifies the nature and application of product(s) he or she wishes to buy, and the seller tries to match these requests almost instantly or in a very short time*. John Deighton stated that interactive marketing features "*the ability to address the individual once more in a way that takes into account his or her unique response*" (Deighton 1996).

Nowadays, with the growth in use of smart electronic devices with communication ability, the recording of customers data and preferences is made easier. Moreover, digital broadcasting in combination with smart television systems, expands the customer's base greatly and enables the process of recording and analyzing feedback data from tele-viewers, resulting in advanced marketing and advertising methods. In addition, the convergence of Interactive Broadcasting Systems and IP Multimedia Subsystems enables even more optimum methods.

New business opportunities are arising in establishing enriched and personal customers relationships, to optimally understand and fulfill their requests. Travel and tourism industry is one of the most progressive industries in the utilization of information and communication technologies (ICT), providing an ideal field to investigate the influence of sophisticated Customer Relationship Management systems (CRM) in marketing. E-marketing systems enable the efficient collection and analyses of customers data and revealing their preferences to e-marketers mining in order to rapidly respond or even better predict the customers' needs.

In addition, the importance of Social Media as effective interactive marketing has significantly increased. In fact, the evolution of Web 2.0 and social media offer a low cost but very effective global marketing platform with a variety of applications suitable for effective destination marketing. Destination Management Organizations (DMOs) can take advantage of the popularity of these tools and reach a global audience with limited resources. Especially when their funding from public sector has decreased due to budget cuts, it is required to achieve higher *value-for-money* efficiency in marketing budgets.

The secondary trading of spectrum in the new digital broadcasting era is an interesting application of Interactive Marketing. Cognitive Radio technology enables dynamic sharing of licensed spectrum with unlicensed customers through real-time spectrum markets. Assuming in a liberalized exploitation of TV White Spaces, new business strategies and marketing opportunities have arisen based on "*Spectrum Commons*" and *"Spectrum Markets"* regimes [3], [47].

## VII. OPTIMIZATION

*Optimization* is defined *as an act, process, or methodology of making something (as a design, system, or decision) as fully perfect, functional, or effective as possible.* As far as Computer Science is concerned, optimization describes the efforts to *increase the speed and efficiency of an algorithm/mechanism/program, by rewriting instructions*. In this section several optimization methods, algorithms and mechanisms are presented, exploiting existing protocols and mechanisms in wireless networking.

In recent years, with the growth in use of mobile devices, there has been a great demand on wireless networks and applications that rely on wireless networks in order to support users' mobility. To meet such demands, the wireless technology of ad-hoc networks has emerged. Wireless Ad-hoc Networks (WANets) are based on decentralized architecture without any pre-existing infrastructure like wired routers or wireless access points but each node participates in routing by forwarding data to other nodes. Self-organization mechanisms are used to make the planning, configuration, management, optimization and healing of such networks simpler and faster.

Examples of such networks are the Mobile Ad-hoc Networks (MANETs), Wireless Sensor Networks (WSNs), Peer-to-Peer (P2P) and Vehicle-to-Vehicle (V2V) networks. Each device of network is free to move anywhere inside the network causing often changes in its relationships with other devices and in network's topology. Therefore, such networks have several interesting characteristics and challenges for optimization. Adaptation methods are used from nodes to adjust themselves into environment conditions [48].

Social mobility is used by some schemes for optimization of End-to-End resources' availability and nodes' survivability. The important nodes, tagged by *Social Centrality* mechanisms, are being offloaded via outsourcing collaborative strategies in order to achieve energy conservation and better performance [49], [50], [51], [52].

Broadcasting is a common operation in MANETs but as it usually uses the flooding technique, it may cause the broadcast storm problem affecting network's performance and reachability. By using adaptation methods, a Counter-based broadcasting scheme addresses this problem in Ref. [53].

Vehicles have much more power than mobile devices used in other types of ad-hoc networks. Reliability and End-to-End reachability are two significant challenges in Vehicle-to-Vehicle (V2V) networks and Vehicle Ad-Hoc Networks (VANETs), determined by specifying reliable mobility models and routing protocols. Since in V2V networks the nodes are moving vehicles, the network's topology is rapidly changing and rarely allows a connected path between the source node and the destination. For the establishment and connectivity of these networks, cars can be used as devices/nodes, establishing Vehicular Ad-Hoc Networks (VANETs). Network size assumption is one key for designing an efficient routing protocol [54]. Exploiting the movement synchronization is another key for optimization of file transmissions and file-sharing services [55], [56]. Movement synchronization schemes that are using the Message Ferry (MF) mobile Peer in Mobile Peer-to-Peer (MP2P) and Vehicular Peer-to-Peer (VP2P) systems, can be found in Ref. [57], [58]. In order to achieve higher resource availability and efficiency in nodes seeding, a hybridized Bittorrent protocol may be used [59]. Moreover, an adaptive Backoff Algorithm is proposed [60] for enhancing the data delivery ratio.

## VIII. CONCLUSION

Technological evolution has lead to a the great demand for mobility and connectivity with the development of the pioneer and promising sector of Wireless Networks playing a major role in this field. In the previous sections, existing technological and scientific approaches are presented, introducing several issues and challenges in important topics of wireless networks. A variety of innovative algorithms with efficient mechanisms and protocols are proposed and implemented for addressing these challenges. In addition, optimization methods are described, enhancing existing techniques and protocols.